# Cosmic Error and
# Statistics of Large Scale Structure

István Szapudi[1] and Stéphane Colombi[1,2]

[1]NASA/Fermilab Astrophysics Center, Fermi National Accelerator Laboratory, Batavia, IL 60510-0500
[2]CITA, 60 St George St., Toronto, ON, Canada, M5S 1A7

## Abstract

We use a generating function approach to examine the errors on quantities related to counts in cells extracted from galaxy surveys.

The measurement error, related to the finite number of sampling cells, is disentangled from the "cosmic error", due to the finiteness of the survey. Using the hierarchical model and assuming locally Poisson behavior, we identified three contributions to the cosmic error:

- ***The finite volume effect*** is proportional to the average of the two-point correlation function over the whole survey. It accounts for possible fluctuations of the density field at scales larger than the sample size.
- The ***edge effect*** is related to the geometry of the survey. It accounts for the fact that objects near the boundary carry less statistical weight than those further away from it.
- The ***discreteness effect*** is due to the fact that the underlying smooth random field is sampled with finite number of objects. This is the "shot noise" error.

To check the validity of our results, we measured the factorial moments of order $N \leq 4$ in a large number of small subsamples randomly extracted from a hierarchical sample realized by Raighley-Lévy random walks. The measured statistical errors are in excellent agreement with our predictions. The probability distribution of errors is increasingly skewed when the order $N$ and/or the cell size increases. This suggests that "cosmic errors" tend to be systematic: it is likely to underestimate the true value of the the factorial moments.

Our study of the various regimes showed that the errors strongly depend on the clustering of the system, i.e., on the hierarchy of underlying correlations. The Gaussian approximation is valid only in the weakly non-linear regime, otherwise it severely underestimates the true errors.

We study the concept of "number of statistically independent cells" (re)defined as the number of sampling cells required to have the measurement error of same order as the cosmic error. This number is found to depend highly on the statistical object under study and is generally quite different from the number of cells needed to cover the survey volume. In light of these findings, we advocate high oversampling for measurements of counts in cells.

## 1. Introduction

The distribution of galaxies is generally admitted to be homogeneous at scales larger than $\sim 150$ Mpc. Statistics naturally characterizes this distribution under the assumption that galaxies are a discrete realization of a continuous random field. This underlying smooth field is related to the distribution of luminous matter, thus it might not necessarily represent the total mass contained in the Universe. The primary purpose of quantifying structure in galaxy catalogs is to measure the properties of the underlying random field. Practically, however, galaxy surveys always cover a finite volume of space and contain finite number of objects. Possible fluctuations of the random field at the boundaries of the survey and at scales larger than the survey size, together with the Poisson fluctuations related to the discrete nature of the sample, introduce uncertainties on the measurements. These effects are present in any finite galaxy catalog, hence the name "cosmic error". This paper puts forth a quantitative analysis of the situation by calculating the theoretically expected errors on statistics related to counts in cells.

Any statistic, aiming at extracting the properties of the underlying random field, measures deviations from a homogeneous random distribution, i.e., one without any correlation between galaxies. The most widely used measures are the two-point correlation function $\xi_2$, and its fourier transform the power-spectrum $\langle|\delta_k|^2\rangle$ (see, e.g., Peebles 1980). Function $\xi_2$ corresponds to the excess probability of pairs compared to random. Uncertainties on the measurement of $\xi_2$ and of $\langle|\delta_k|^2\rangle$ have been discussed by various authors (e.g., Peebles 1973; Peebles 1980; Landy & Szalay 1992; Hamilton 1993; Bernstein 1994; Feldman, Kaiser, & Peacock 1994; Colombi, Bouchet & Schaeffer 1995, hereafter CBSII) and, as a result, the correlation function has become a well controlled tool. However, the large degree of inhomogeneity in the galaxy distribution, manifesting in large voids (e.g., de Lapparent *et al.* 1986; Kirshner *et al.* 1987; Geller & Huchra 1989), clusters and superclusters (e.g., Abell 1958; Bahcall 1988) is not fully described by this statistic, which only accounts for a Gaussian distribution adequately. To measure non-Gaussian features, higher order correlation functions, $\xi_N$, are needed (e.g., Peebles & Groth 1975; Groth & Peebles 1977; Fry & Peebles 1978; Sharp, Bonometto, & Lucchin 1984). Unfortunately it is difficult to measure and interpret the $N$-point correlation functions, especially when $N \geq 5$, mostly because of the large number of parameters involved. In particular, the expected uncertainties on the measurements are rather difficult to estimate.

Alternative to correlation functions, counts in cells estimate the probability of finding $N$ objects in a cell of given size thrown at random in the survey. They depend on integrals of the $M$-point correlation functions, $M \geq 2$, thus characterize, although indirectly, the clustering of galaxies to greater accuracy than the two-point function. Describing only the scaling of the underlying distribution with the cell size, they are much simpler to deal with than the $N$-point correlation functions. Methods based on counts in cells and related statistics such as moments, and moment correlators were thus applied to several galaxy catalogs (e.g., Alimi, Blanchard & Schaeffer 1990; Maurogordato, Schaeffer & da Costa 1992; Szapudi, Szalay & Boschán 1992; Meiksin, Szapudi & Szalay 1992; Bouchet *et al.* 1993; Gaztañaga 1994; Szapudi *et al.* 1995) and $N$-body simulations (e.g., Bouchet, Schaeffer & Davis 1991; Bouchet & Hernquist 1992; Lucchin *et al.* 1994; Baugh, Gaztañaga & Efstathiou 1995; Colombi, Bouchet, & Hernquist 1995). The assessment of the errors on these measurements is even more delicate than the measurements themselves.

One possible procedure to estimate the errors consists of generating a large number of random realizations modeling the data set, and measure the dispersion of the measurements experimentally (see, e.g., Baugh, Gaztañaga & Efstathiou 1995). Such Monte Carlo methods are rather costly, since to date the main technique to generate an artificial sample with realistic statistics is the costly $N$-body simulation. Therefore the number of realizations used for such measurements is severely limited by the available computer resources. The other option is a full scale analytic calculation. So far, no real detailed analytic study of the errors on counts in cells has been done, except for the void probability (CBSII), and to some extent for the second order moment (which is more or less equivalent to the two-point correlation function, e.g., Efstathiou *et al.* 1990; CBSII). Our aim here is thus to deal in detail with the errors on quantities related to counts in cells, especially the factorial moments of the count probability distribution. These latter can be simply calculated from the counts in cells, and they estimate the moments of the underlying smooth random field in a consistent and unbiased manner (e.g., Szapudi & Szalay 1993a, hereafter SSI).

As will be shown later, the measurement of the probability distribution of counts in cells in a galaxy catalog is burdened with errors from various possible sources (if all systematics of the observations are discounted):

- The usual way of estimating probability distributions consists of building histograms from finite number of randomly thrown cells, which introduces *measurement errors*.
- The survey spans only a finite portion of the Universe, which introduces *finite volume error*, related to fluctuations of the density field at scales larger than the sample size.
- The geometry of the survey causes *edge effects*, due to the fact that objects near the edge of the survey are given less statistical weight than objects far away.
- The galaxies sample the underlying continuous field with finite number of objects, which creates *discreteness errors*.

The first of these contributions can in principle be eliminated by efficient computer algorithms (e.g., Szapudi 1995). The other three are, however, present in all samples even under ideal conditions, because they are simply due to the finiteness of the part of the Universe we have access to. These errors can be *systematic*, i.e. the mean can be substantially different from the most likely value. As will be shown later using a Rayleigh-Lévy hierarchical sample, it is more likely to *underestimate* the real moments of the distribution than to overestimate them (see also Colombi, Bouchet, & Schaeffer 1994, hereafter CBSI; Colombi, Bouchet, & Hernquist 1995).

The presentation is organized as follows: in §2, after introducing the general formalism, the measurement errors are evaluated. §3 describes the framework of the hierarchical model, which is, together with a locally Poisson approximation, used in §4 to calculate the errors on factorial moments of order up to four explicitly. In §5, the theoretical results are compared to error estimates from subsamples extracted from a Raighley-Lévy fractal. The study of the distribution of errors shows that it can be strongly skewed implying that the errors are systematic. In §6, we summarize the results and discuss some applications, such as the validity of the Gaussian approximation and the concept of "number of statistically independent cells". The appendices contain mathematical derivations which would have interrupted the flow of the main text.

## 2. General Formalism

In this section, we present a formalism for calculating the theoretically expected errors on estimates related to counts in cells measured in a finite galaxy catalog. Note that the following formalism can be applied to any random distribution of points windowed by a finite box.

Let us imagine that we have a galaxy catalog of volume $V$, corresponding to length scale $L$, and we attempt to measure the probability distribution in cells of volume $v$, corresponding to length scale $\ell$. The theoretical dispersion of the measurement is partly due to throwing finite number of cells and partly reflects the finite nature of the data set. The former effect can in principle be eliminated by throwing a very large (or infinite) number of cells. The latter effect corresponds to the "cosmic error" in our data set.

Let $P_N$ denote the probability of finding $N$ galaxies in a cell of size $\ell$. The factorial moments (see, e.g., SSI) are defined as

$$F_k = \langle (N)_k \rangle = \sum (N)_k P_N, \tag{1}$$

where $(N)_k = N(N-1)...(N-k+1)$ is the $k$-th falling factorial of $N$. The ensemble average $\langle\ \rangle$ can be evaluated through the probability distribution $P_N$ for any 1-point quantity. We introduce the generating function of the probability distribution

$$P(x) = \sum_{N=0}^{\infty} P_N x^N, \tag{2}$$

and $F(x) = P(x+1)$ (see, e.g., SSI), the exponential generating function of the factorial moments, i.e.

$$F(x) = \sum_{k\geq 0} F_k \frac{x^k}{k!}. \tag{3}$$

Similarly to the above definitions we introduce the quantities $\tilde{P}_N^C$, $\tilde{F}_k^C$ and the generating functions $\tilde{P}^C(x)$ and $\tilde{F}^C(x)$, corresponding respectively to the *estimates* of $P_N$, $F_k$, $P(x)$ and $F(x)$ from randomly throwing $C$ number of cells in the galaxy catalog. Note that this notation refers implicitly to a particular set of cells; another set of cells can give different results. The equation

$$\tilde{F}^C(x) = \tilde{P}^C(x+1) \tag{4}$$

still holds for a given set of cells, so it is quite easy to pass from $\tilde{P}^C(x)$ to $\tilde{F}^C(x)$. If $N_i$ denotes the number of objects in cell "$i$", then the following equations are true

$$\begin{aligned} \tilde{P}_N^C &= \frac{1}{C}\sum_{i=1}^{C} \delta(N_i = N) \\ \tilde{P}^C(x) &= \sum_{N\geq 0} x^N \tilde{P}_N^C = \frac{1}{C}\sum_{i=1}^{C} x^{N_i}, \end{aligned} \tag{5}$$

and $\delta(N = M)$ is the Kronecker-delta. It is easy to see that the ensemble average of $\tilde{P}^C(x)$ is $\left\langle \tilde{P}^C(x) \right\rangle = P(x)$, the underlying generating function. The usual measure of error on the counts in cells and the factorial moments is the dispersion

$$\begin{aligned} \left(\Delta \tilde{P}_N\right)^2 &= \left\langle \left\langle (\tilde{P}_N^C)^2 \right\rangle \right\rangle_C - \left\langle \left\langle \tilde{P}_N^C \right\rangle \right\rangle_C^2 \\ \left(\Delta \tilde{F}_k\right)^2 &= \left\langle \left\langle (\tilde{F}_k^C)^2 \right\rangle \right\rangle_C - \left\langle \left\langle \tilde{F}_k^C \right\rangle \right\rangle_C^2, \end{aligned} \tag{6}$$

where the ensemble average is taken first over the measurements, and the operator $\langle\ \rangle_C$ averages over all possible ways of throwing $C$ cells in the survey volume $V$. The error generating function $E^{C,V}(x,y)$ can be defined in such a way that the coefficients of $(xy)^N$ in its series expansion are $\left(\Delta\tilde{P}_N\right)^2$:

$$E^{C,V}(x,y) = \sum_{N,M} \left[\left\langle\left\langle \tilde{P}^C_N \tilde{P}^C_M \right\rangle\right\rangle_C - \left\langle\left\langle \tilde{P}^C_N \right\rangle\right\rangle_C \left\langle\left\langle \tilde{P}^C_M \right\rangle\right\rangle_C\right] x^N y^M. \tag{7}$$

Similarly, the exponential generating function for the errors on factorial moments is

$$E^{C,V}(x+1,y+1) = \sum_{N,M} \left[\left\langle\left\langle \tilde{F}^C_N \tilde{F}^C_M \right\rangle\right\rangle_C - \left\langle\left\langle \tilde{F}^C_N \right\rangle\right\rangle_C \left\langle\left\langle \tilde{F}^C_M \right\rangle\right\rangle_C\right] \frac{x^N y^M}{N!M!}. \tag{8}$$

Straightforward calculation yields

$$E^{C,V}(x,y) = \left\langle\left\langle \tilde{P}^C(x)\tilde{P}^C(y) \right\rangle\right\rangle_C - \left\langle\left\langle \tilde{P}^C(x) \right\rangle\right\rangle_C \left\langle\left\langle \tilde{P}^C(y) \right\rangle\right\rangle_C. \tag{9}$$

Clearly $\left\langle\left\langle \tilde{P}^C(x) \right\rangle\right\rangle_C = P(x)$. From equation (5), we have

$$\begin{aligned} \left\langle\left\langle \tilde{P}^C(x)\tilde{P}^C(y) \right\rangle\right\rangle_C &= \frac{1}{C^2}\left\langle \sum_{i,j} \left\langle x^{N_i} y^{N_j} \right\rangle \right\rangle_C \\ &= \frac{1}{C^2}\left\langle \sum_{i=1}^{C} \left\langle (xy)^{N_i} \right\rangle + \sum_{i\neq j}^{C} \left\langle x^{N_i} y^{N_j} \right\rangle \right\rangle_C \\ &= \frac{P(xy)}{C} + \left(1-\frac{1}{C}\right)\left\langle \tilde{P}^\infty(x)\tilde{P}^\infty(y) \right\rangle, \end{aligned} \tag{10}$$

We introduced the notation

$$\left\langle \tilde{P}^\infty(x)\tilde{P}^\infty(y) \right\rangle \equiv \frac{1}{\hat{V}^2}\int_{\hat{V}} d^D r_1 d^D r_2 P(x,y), \tag{11}$$

where $D$ is the dimension of the survey, $P(x,y)$ is the generating function of the underlying bivariate probability distribution $P_{NM}$ for two cells located at $r_1$ and $r_2$ with $N$ and $M$ particles respectively. The integral is performed over the volume $\hat{V}(\ell)$ corresponding to all possible positions of cells entirely included in the survey volume $V$. Therefore, we have $\hat{V} \simeq V$ in the case $v \ll V$, and $\hat{V} \ll V$ when the cell size becomes comparable to the survey size. In the second line of equation (10), the sum is separated into two parts $i = j$ and $i \neq j$. The second part proves to be the Monte Carlo realization of equation (11). In what follows, we drop the index $\infty$ from $\tilde{P}^\infty(x) = \tilde{P}(x)$ for simplicity. If the volume $V$ tends to infinity in equation (11), "statistically independent" cells will dominate the ensemble averaging, therefore $\left\langle \tilde{P}(x)\tilde{P}(y) \right\rangle \to P(x)P(y)$, and the error generating function is

$$E^{C,\infty}(x,y) = \frac{P(xy) - P(x)P(y)}{C}. \tag{12}$$

With the notation

$$E^{\infty,V}(x,y) = \left\langle \tilde{P}(x)\tilde{P}(y) \right\rangle - \left\langle \tilde{P}(x) \right\rangle \left\langle \tilde{P}(y) \right\rangle, \tag{13}$$

we have the equation

$$E^{C,V}(x,y) = \left(1-\frac{1}{C}\right) E^{\infty,V}(x,y) + E^{C,\infty}(x,y). \tag{14}$$

The total error on the measurement of factorial moments can thus be (approximately) disentangled into two parts: the "cosmic error", $E^{\infty,V}$, due to the finiteness of the sample, and the measurement error $E^{C,\infty}$ due to using only $C$ cells.

This intuitive notion was present in the literature before (see, e.g., Hamilton 1985; Politzer & Preskill 1986; Maurogordato & Lachièze-Rey 1987), and materialized in the fuzzy concept of "number of statistically independent cells". This number, $C^*$, corresponds to the number of sampling cells needed to extract all (actually most of) the statistically relevant information in the survey. From equation (14), it is natural to choose $C^*$ such that the measurement error and the cosmic error are of the same order. Note, however, that some residual information still can be extracted using more cells, since the measurement error can be rendered arbitrarily small. Our choice of $C^*$ qualitatively matches the calculations of Politzer & Preskill (1986) at least for the void probability distribution function (CBSII). Another simple choice often found in the literature is to cover the sampled volume uniformly by cells requiring $C^* \simeq V/v \equiv C_V$. We shall see in § 6 that with our more natural definition, $C^*$ depends sensitively on the statistical object under study and can be different from $C_V$ by several orders of magnitude. Note that in any case, to reduce the errors as much as possible, it is advisable to use many cells to measure count in cells, so that the measurement error is negligible compared to the cosmic error.

In the following, we will evaluate $E^{\infty,V}(x,y)$, the generating function of the cosmic error. Subsequently by "errors" we mean cosmic errors since the measurement errors are avoidable in principle.

Following CBSII, the integration in equation (11) is split into two parts according to whether the cells overlap or not:

$$\left\langle \tilde{P}(x)\tilde{P}(y)\right\rangle = \left\langle \tilde{P}(x)\tilde{P}(y)\right\rangle_{\rm overlap} + \left\langle \tilde{P}(x)\tilde{P}(y)\right\rangle_{\rm disjoint} \tag{15}$$

with

$$\left\langle \tilde{P}(x)\tilde{P}(y)\right\rangle_{\rm overlap} \equiv \frac{1}{\hat{V}^2}\int_{r\leq 2\ell} d^D r_1 d^D r_2 P(x,y), \tag{16}$$

$$\left\langle \tilde{P}(x)\tilde{P}(y)\right\rangle_{\rm disjoint} \equiv \frac{1}{\hat{V}^2}\int_{r\geq 2\ell} d^D r_1 d^D r_2 P(x,y), \tag{17}$$

and $r = |r_1 - r_2|$. As shown below, the term $\left\langle \tilde{P}(x)\tilde{P}(y)\right\rangle_{\rm overlap}$ contains two contributions: the *discreteness effect* brought by the sampling of the underlying smooth distribution with a finite number of points, and the *edge effect* due to the fact the statistical weight given to points decreases toward the boundary of the catalog (e.g., Ripley 1988). The term $\left\langle \tilde{P}(x)\tilde{P}(y)\right\rangle_{\rm disjoint}$, shown later to be proportional to the average of the correlation function over $\hat{V}$, is due to fluctuations of the underlying random field at scales larger than the sample size: it is related to the *finite volume* of the catalog.

The calculation of $\left\langle \tilde{P}(x)\tilde{P}(y)\right\rangle_{\rm overlap}$ involves the generating function $P(x,y)$ for overlapping cells. As proved in Appendix A., $P(x,y)$ can be obtained from the the trivariate probability generating function of the three non-overlapping cells corresponding to the two original cells as

$$P(x,y) = P(x,xy,y), \tag{18}$$

where the parameters $xy$, $x$, and $y$ are associated with the cell formed by the overlap area, and the rest of each cell ($I$,$H$, and $J$ on Fig. 1 respectively).

According to the previous equations, error calculations on statistics related to counts in cells in a finite survey involve generating functions for counts in disjoint cells up to trivariate level. These functions can be expressed as integrals over the $N$-point correlation functions $\xi_N(r_1,\ldots,r_N)$ (Balian & Schaeffer 1989, hereafter BS, SSI, Szapudi & Szalay 1993b, hereafter SSII), i.e.,

$$P(x) = \exp\left\{\sum_{N=1}^{\infty} \frac{\bar{N}^N (x-1)^N}{N! v^N} \int_v d^D r_1 \ldots d^D r_N \xi_N(r_1,\ldots,r_N)\right\}, \tag{19}$$

$$\begin{aligned} P(x,y) &= \exp\left\{\sum_{N,M} \frac{(x-1)^N (y-1)^M \bar{N}^{N+M}}{N! M! v^{N+M}} \right. \\ &\left. \int_{v_1} d^D r_1 \ldots d^D r_N \int_{v_2} d^D r_{N+1} \ldots d^D r_{N+M} \xi_{N+M}(r_1,\ldots,r_{N+M})\right\}, \end{aligned} \tag{20}$$

and

$$\begin{aligned} P(x,y,z) &= \exp\left\{\sum_{N,M,S} \frac{(x-1)^N (y-1)^M (z-1)^S \bar{N}^{N+M+S}}{N! M! S! v^{N+M+S}} \right. \\ & \int_{v_1} d^D r_1 \ldots d^D r_N \int_{v_2} d^D r_{N+1} \ldots d^D r_{N+M} \\ &\left. \int_{v_3} d^D r_{N+M+1} \ldots d^D r_{N+M+S} \xi_{N+M+S}(r_1,\ldots,r_{N+M+S})\right\}. \end{aligned} \tag{21}$$

The quantity $\bar{N} \equiv \sum N P_N = F_1$ is the average number of objects per cell. These equations together with equations (14), (15) and (18) solve the problem in principle: a model for the integrals of the higher order correlation functions yields the errors on factorial moments as a set of finite, although complicated, expressions.

## 3. Hierarchical Model

In the highly nonlinear regime $\xi_2 \gg 1$, an ansatz for the structure of the $N$-point correlation functions is the *hierarchical model* (e.g., Peebles 1980; BS)

$$\xi_N(r_1,\ldots,r_N) = \sum_{k=1}^{K(N)} Q_{Nk} \sum^{B_{Nk}} \prod^{N-1} \xi(r_{ij}), \tag{22}$$

where $\xi(r) \equiv \xi_2(r) = (r/r_0)^{\gamma}$. Such a model seems to give a reasonably good description of the statistics measured in the galaxy distribution (e.g., Groth & Peebles 1977; Fry & Peebles 1978; Sharp, Bonometto, & Lucchin 1984; Szapudi, Szalay & Boschán 1992; Meiksin, Szapudi & Szalay 1992; Szapudi *et al.* 1995) and in $N$-body simulations (e.g., Bouchet, Schaeffer & Davis 1991; Bouchet & Hernquist 1992; Fry, Mellott & Shandarin 1993; Bromley 1994; Lucchin *et al.* 1994; CBSI; CBSII; Colombi, Bouchet, & Hernquist 1995), particularly in the nonlinear regime.

In equation (22), the summation is over all possible $N^{N-2}$ trees with $N$ vertices. In the sum, every $\xi(r_{ij})$ corresponds to an edge $r_{ij} = \mid r_i - r_j \mid$ in a tree spanned by $r_1,\ldots,r_N$. For every tree, there is a product of $N-1$ two-point functions, there is a summation over all the $B_{Nk}$ labelings of all the $K(N)$ distinct trees. The scale independent average $Q_N$ is defined as

$$Q_N = \frac{\sum_{k=1}^{K(N)} Q_{Nk} B_{Nk} F_{Nk}}{\sum_{k=1}^{K(N)} B_{Nk} F_{Nk}}, \tag{23}$$

where $F_{Nk}$ are the form factors associated with the shape of cell of size unity (see Boschán, Szapudi & Szalay 1994 for details)

$$F_{Nk} = \int_1 d^3q_1 \dots d^3q_N \prod^{N-1} \left\{ | q_i - q_j |^\gamma \int_1 d^3p_1 d^3p_2 |p_1 - p_2|^{-\gamma} \right\}^{-1}. \tag{24}$$

The above product is running over the $N-1$ edges of a tree. Since the number of all tree graphs with $N$ vertices is $N^{N-2}$, the generating function takes the following form:

$$P(x) = \exp \left\{ \sum_{N=1}^{\infty} (x-1)^N \Gamma_N Q_N \right\}, \tag{25}$$

with $Q_1 \equiv Q_2 \equiv 1$. The following shorthand notation is used

$$\Gamma_N = \frac{N^{N-2} \bar{N}^N \bar{\xi}^{N-1}}{N!}, \tag{26}$$

where $\bar{\xi}$ is the average of the two-point correlation function in a cell

$$\bar{\xi} = v^{-2} \int d^3r_1 d^3r_2 \xi(r_1, r_2). \tag{27}$$

Note that equation (25) is also valid if the $N$-point correlation functions obey the *scaling relation* (BS)

$$\xi_N(\lambda r_1, \dots, \lambda r_N) = \lambda^{-\gamma(N-1)} \xi_N(r_1, \dots, r_N), \tag{28}$$

which is more general than the hierarchical model (22).

To obtain workable expressions of the bivariate and trivariate distributions supplementary assumptions are needed. We quote here two approximations, hereafter SS and BeS, worked out respectively by SSI, SSII and by Bernardeau & Schaeffer (1992).

If the distance $r$ between the two cells is large compared to the cell size $\ell$, the correlation between two particles in each cell is $\sim \xi(r)$. The integral in equation (20) can be then well approximated as $Q_{N+M}\Gamma_N\Gamma_M NM\xi$ (see Szapudi, Szalay & Boschán 1992; Szapudi *et al.* 1995) up to linear order in $\xi/\bar{\xi}$. This was found to be fairly accurate even when the cells are touching. The bivariate generating function can thus be written, in this framework,

$$\begin{aligned} P(x,y) &\simeq P(x)P(y) \exp\{R(x,y)\} \\ &\simeq P(x)P(y)\left[1 + R(x,y)\right] + \mathcal{O}(\xi^2/\bar{\xi}^2). \\ R(x,y) &= \xi \sum_{M=1,N=1}^{\infty} (x-1)^M (y-1)^N Q_{N+M} \Gamma_M \Gamma_N NM. \end{aligned} \tag{29}$$

This is the SS approximation. A similar expression was developed in SSII for the trivariate generating function which could be used, in conjunction with the results of the previous section to evaluate the cosmic errors. However, the locally Poisson approximation developed in the next section eliminates the need for using it, therefore we do not quote it here explicitly.

The other approach, BeS, consists in using a special (but still quite general) case of the hierarchical model. It is assumed that $Q_{Nk}$, the structure constant associated with a tree topology labelled with $k$ can be written as

$$Q_{Nk} = \prod_{i=2}^{\infty} \nu_i^{d_i(k)}, \tag{30}$$

where $\nu_i$ is a weight associated with a vertex with $i$ lines, and $d_i(k)$ is the number of such vertices in tree type $k$. Under this condition, the bivariate generating function is approximated by

$$P(x,y) \simeq P(x)P(y)\left[1 + \tau\{(1-x)\bar{N}\bar{\xi}\}\,\tau\{(1-y)\bar{N}\bar{\xi}\}\,\xi/\bar{\xi}^2\right] + \mathcal{O}(\xi^2/\bar{\xi}^2), \tag{31}$$

where

$$\tau(s) = s\sqrt{2\sum_{N\geq 2}(-s)^{N-2} Q_N \frac{N^{N-2}(N-1)}{N!}}. \tag{32}$$

Again, this approximation can be generalized to higher order multivariate generating functions (see Bernardeau & Schaeffer 1992).

Although the two approximations SS and BeS appear quite different formally, we shall see in § 6 that in the regimes we considered in this paper they are practically identical.

Note, that the hierarchical model does not hold in the weakly non-linear regime $\bar{\xi}_2 \ll 1$ (e.g., Fry 1984; Bernardeau 1992), where similar but different formalism has to be applied. We conjecture that the final result will be quite similar (Bernardeau 1994a), although the proof is left for future work.

## 4. Calculation of the Cosmic Error

### 4.1. Contribution from Disjoint Cells

Using the approximations of the previous section, the finite volume effect (second term in eq. [15]) can be computed.

$$\left\langle \tilde{P}(x)\tilde{P}(y)\right\rangle_{\rm disjoint} \simeq P(x)P(y)\left[1 + \bar{\xi}(\hat{L})\Sigma_{\rm X}(x,y)\right], \tag{33}$$

where

$$\Sigma_{\rm SS}(x,y) = \sum_{M=1,N=1}^{\infty} (x-1)^M (y-1)^N Q_{N+M} \Gamma_M \Gamma_N N M. \tag{34}$$

$$\Sigma_{\rm BeS}(x,y) = \tau\{(1-x)\bar{N}\bar{\xi}\}\,\tau\{(1-y)\bar{N}\bar{\xi}\}\,/\bar{\xi}^2, \tag{35}$$

are derived from the two approximations quoted for the bivariate generating function, and $\bar{\xi}(\hat{L})$ is given by the following integral

$$\bar{\xi}(\hat{L}) \equiv \hat{V}^{-2}\int_{r_1\in\hat{V},r_2\in\hat{V},r\geq 2\ell} d^D r_1 d^D r_2 \xi(r_1,r_2). \tag{36}$$

When $v/V \ll 1$, this quantity is approximately the average of the two-point correlation function over the survey volume:

$$\bar{\xi}(\hat{L}) \simeq \bar{\xi}(L) \equiv V^{-2} \int_V d^D r_1 d^D r_2 \xi(r_1, r_2), \quad v/V \ll 1. \tag{37}$$

### 4.2. Contribution from Overlapping Cells

The calculation of the first term, $\left\langle \tilde{P}(x)\tilde{P}(y)\right\rangle_{\text{overlap}}$ of equation (15) is burdened with more difficulty than the second one, because the trivariate generating function depends on the separation $r$ between the two overlapping cells in a complicated way. The approximations SS and BeS could be generalized even though the cells are touching if we disregard the fact that the three cells formed by the overlap and the remaining parts are nonspherical. We leave this rather cumbersome calculation for subsequent research.

Instead, we worked out a simple approximation, which, as we shall see in § 5 experimentally, provides sufficiently accurate estimate of first integral of equation (15) for practical purposes. This approximation consists of assuming locally Poisson behavior, i.e., that the correlations inside the union $C_\cup$ of two overlapping cells $C_1$ and $C_2$ are smeared out. A natural consequence of this is that the effects of the nonsphericity of $C_\cup$ can be neglected as well.

Let us denote the volume of $C_\cup$ by $v_\cup$ and the radius of the corresponding spherical cell with the same volume by $\ell_\cup$. For three dimensions, we have (CBSII)

$$\frac{4}{3}\pi\ell_\cup^3 = v_\cup = \frac{4}{3}\pi\ell^3\left[1 + f_3\left(\frac{r}{\ell}\right)\right], \quad f_3(\psi) = \frac{3}{4}\psi - \frac{1}{16}\psi^3, \tag{38}$$

and the two-dimensional case gives

$$\pi\ell_\cup^2 = v_\cup = \pi\ell^2\left[1 + f_2\left(\frac{r}{\ell}\right)\right], \; f_2(\psi) = 1 - \frac{1}{\pi}\left[2\arccos\frac{\psi}{2} - \sqrt{1 - \frac{\psi^2}{4}}\psi\right]. \tag{39}$$

According to the locally Poisson ansatz, the probability of an object to be in a portion of $C_\cup$ is proportional to the volume of this portion. An object can belong to the overlapping part $C_\cap \equiv C_1 \cap C_2$ or the rest $C_i \setminus C_\cap$ of one of the cells with probabilities $p = [1 - f_D(\psi)]/[1 + f_D(\psi)]$ and $q = f_D(\psi)/[1 + f_D(\psi)]$, respectively. The probability $P^S_{H,I,J}$ of finding $H$, $I$ and $J$ objects in $C_1 \setminus C_\cap$, $C_\cap$ and $C_2 \setminus C_\cap$, respectively, under the constraint $H + I + J = S$ (Fig. 1), is a "trinomial" distribution

$$P^S_{H,I,J} = \frac{S!}{H!I!J!} q^{H+J} p^I, \tag{40}$$

with the following generating function

$$P^S(x, y, z) = (qx + py + qz)^S. \tag{41}$$

Since the effects of the nonsphericity of the volume formed by $C_1 \cup C_2$ can be neglected under the locally Poisson assumption, the unconstrained probability $P_{H,I,J}$ of finding $I$, $J$, $K$ respectively in $C_1 \setminus C_\cap$, $C_\cap$ and $C_2 \setminus C_\cap$ is simply

$$P_{H,I,J} = P^\cup_{H+I+J} P^{H+I+J}_{H,I,J}, \tag{42}$$

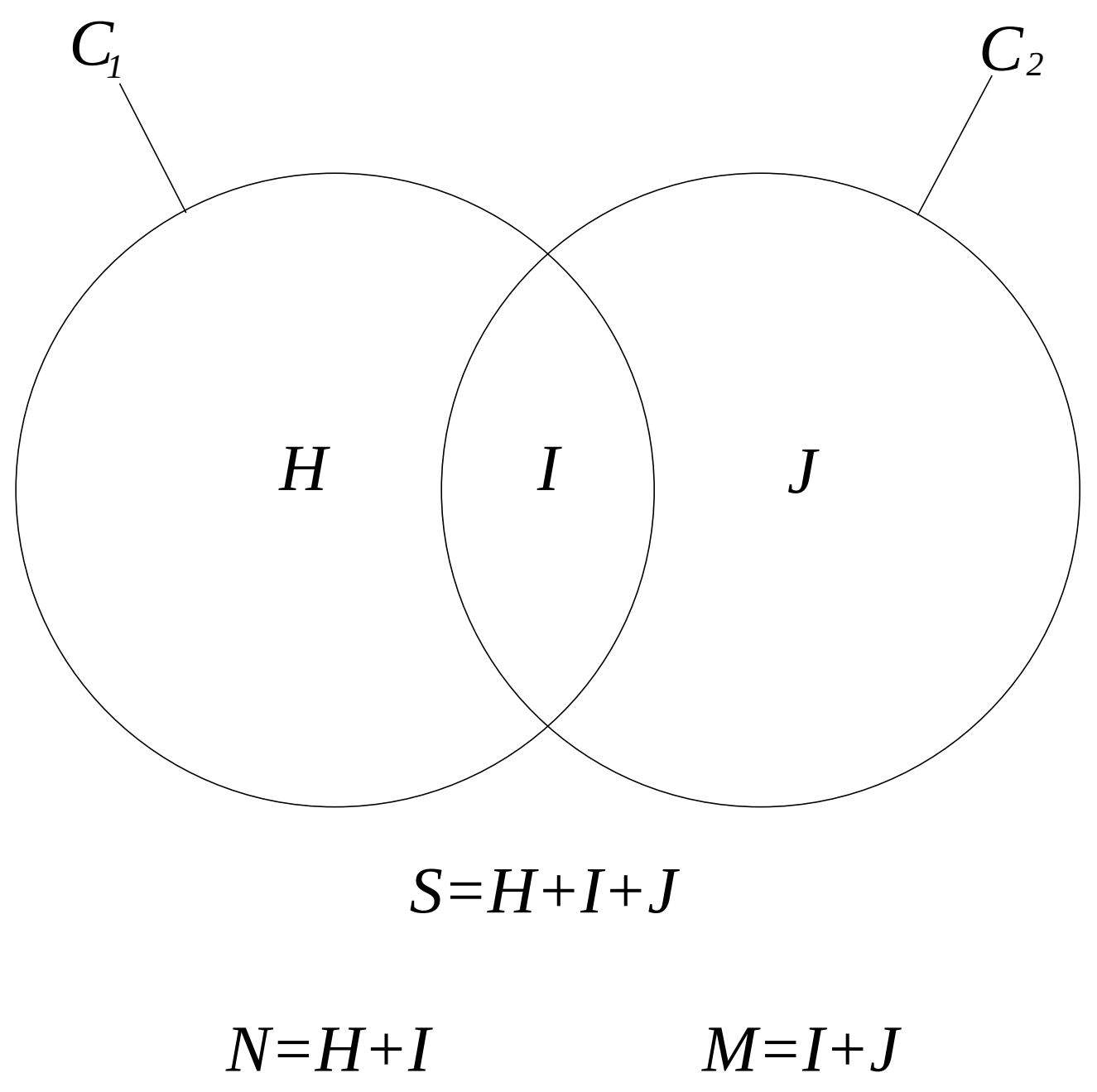

Fig. 1.— This is a symbolic drawing of the two overlapping cells $C_1$ and $C_2$. The intersection $C_\cap \equiv C_1 \cap C_2$ contains $I$ objects. Each remaining part, $C_1 \setminus C_\cap$ and $C_2 \setminus C_\cap$, respectively, contains $H$ and $J$ objects. We compute the probability of having $N = H + I$ objects in $C_1$ and $M = I + J$ objects in $C_2$. The locally Poisson behavior allows us to neglect both the correlations inside the union $C_\cup \equiv C_1 \cup C_2$ of the two cells and the nonsphericity of $C_\cup$.

where the probability of finding $S$ objects in a spherical cell of radius $\ell_{\cup}$ is

$$P_S^{\cup} \equiv P_S|_{\ell=\ell_{\cup}}. \tag{43}$$

The trivariate generating function thus can be written as

$$P(x,z,y) = \sum_S P_S^{\cup}(qx+py+qz)^S = P^{\cup}(qx+py+qz). \tag{44}$$

Then, according to equation (18), the generating function of the bivariate probability $P_{N,M}$ of finding $N = H + I$ and $M = I + J$ in the overlapping cells $C_1$ and $C_2$, respectively, is

$$P^{\flat}(x,y) = P^{\cup}\left(q(x+y)+pxy\right), \tag{45}$$

which takes the following form from equation (25)

$$\begin{aligned} P^{\flat}(x,y) \quad \equiv \quad & \exp\left\{\sum_{N\geq 1} \Gamma_N Q_N \left[1+f_D(\psi)\right]^{-(N-1)\left(\frac{\gamma_3+D-3}{D}\right)}\right. \\ & \left.\left[f_D(\psi)(x+y)+\left[1-f_D(\psi)\right]xy-1-f_D(\psi)\right]^N\right\}, \end{aligned} \tag{46}$$

where $\gamma_3$ is the slope of the three-dimensional correlation function. In this equation, the effect of higher order clustering is fully contained in the term $\Gamma_N Q_N$, evaluated at the original cell size $\ell$ of $C_1$ and $C_2$.

Changing variables $\psi = r/\ell$, the first term of equation (15) becomes

$$\left\langle \tilde{P}(x)\tilde{P}(y)\right\rangle_{\text{overlap}} \simeq \frac{v}{V}\int_0^2 D\psi^{(D-1)}d\psi P^{\flat}(x,y) \tag{47}$$

up to leading order in $v/V$ (see Appendix B. for details). This contribution is inversely proportional to $C_V = V/v$, the number of cells needed to cover the survey volume.

### 4.3. Theoretical Results: Cosmic Error on Factorial Moments

After carrying out the appropriate subtractions, the previous results can be summarized as

$$E^{\infty,V}(x,y) \simeq \bar{\xi}(L)\Sigma_{\rm X}(x,y) + \frac{v}{V}\left\{\int_0^2 D\psi^{(D-1)}d\psi P^{\flat}(x,y) - 2^D P(x)P(y)\right\}, \tag{48}$$

to leading order in $v/V$, where $\Sigma_{\rm X}(x,y)$ is given by equations (34) or (35), $P^{\flat}(x,y)$ by equation (46) and $P(x)$ by equation (25). The cosmic error $\Delta^{\infty,V}\tilde{F}_k$ on the factorial moment of order $k$, can be computed by expanding the generating function $E^{\infty,V}(x+1,y+1)$ and then integrating numerically the function $P^{\flat}$. The final result can be expressed as a function of the $F_j$'s with $j \leq 2k$.

We computed $\Delta^{\infty,V}\tilde{F}_k$ for $k \leq 4$. In the three-dimensional case, $D = 3$, we considered various values of $\gamma$: $\gamma = 1.8$, 1.5, 1.2, and 0.9. In the two-dimensional case, $D = 2$, we carried out the calculation only for $\gamma = \gamma_3 - 1 = 0.8$.

Here we present the results explicitly for $\gamma_3 = 1.8$, $D = 2,3$ and $k \leq 3$. Because of their physical significance, we disentangled the terms corresponding to the finite volume, edge and discreteness effects:

$$\left(\Delta^{\infty,V}\tilde{F}_k\right)^2 = \left(\Delta^{\rm finite}\tilde{F}_k\right)^2 + \left(\Delta^{\rm edge}\tilde{F}_k\right)^2 + \left(\Delta^{\rm discrete}\tilde{F}_k\right)^2. \tag{49}$$

Up to leading order in $v/V$ the cosmic error on the first moment (average count) is

$$\left(\Delta^{\rm finite}\tilde{F}_1\right)^2_{\rm SS,BeS} = \bar{N}^2\bar{\xi}(L), \quad (50)$$

$$\left(\Delta^{\rm edge}\tilde{F}_1\right)^2_{D=2} = 6.431\bar{N}^2\bar{\xi}\frac{v}{V}, \quad (51)$$

$$\left(\Delta^{\rm edge}\tilde{F}_1\right)^2_{D=3} = 5.508\bar{N}^2\bar{\xi}\frac{v}{V}, \quad (52)$$

$$\left(\Delta^{\rm discrete}\tilde{F}_1\right)^2_{D=2,3} = \bar{N}\frac{v}{V}. \quad (53)$$

For the second factorial moment

$$\left(\Delta^{\rm finite}\tilde{F}_2\right)^2_{\rm SS} = 4\left(1+2\bar{\xi}\,Q_3+\bar{\xi}^2\,Q_4\right)\bar{N}^4\bar{\xi}(L), \quad (54)$$

$$\left(\Delta^{\rm finite}\tilde{F}_2\right)^2_{\rm BeS} = 4\left(1+2\bar{\xi}\,Q_3+\bar{\xi}^2\,Q_3^2\right)\bar{N}^4\,\bar{\xi}(L), \quad (55)$$

$$\left(\Delta^{\rm edge}\tilde{F}_2\right)^2_{D=2} = \left(11.29+3.781\,\bar{\xi}+31.12\,\bar{\xi}\,Q_3+33.60\,\bar{\xi}^2\,Q_4\right)\bar{N}^4\,\bar{\xi}\frac{v}{V} \quad (56)$$

$$\left(\Delta^{\rm edge}\tilde{F}_2\right)^2_{D=3} = \left(17.05+3.417\,\bar{\xi}+45.67\,\bar{\xi}\,Q_3+42.24\,\bar{\xi}^2\,Q_4\right)\bar{N}^4\,\bar{\xi}\frac{v}{V} \quad (57)$$

$$\left(\Delta^{\rm discrete}\tilde{F}_2\right)^2_{D=2} = \left(0.919+4\,\bar{N}+0.803\,\bar{\xi}+10.16\,\bar{N}\,\bar{\xi}+8.642\,\bar{N}^2\,\bar{\xi}^2\,Q_3\right)\bar{N}^2\frac{v}{V} \quad (58)$$

$$\left(\Delta^{\rm discrete}\tilde{F}_2\right)^2_{D=3} = \left(0.648+4\,\bar{N}+0.502\,\bar{\xi}+8.871\,\bar{N}\,\bar{\xi}+6.598\,\bar{N}^2\,\bar{\xi}^2\,Q_3\right)\bar{N}^2\frac{v}{V}. \quad (59)$$

Finally, for the third factorial moment

$$\left(\Delta^{\rm finite}\tilde{F}_3\right)^2_{\rm SS} = 9\left(1+2\,\bar{\xi}+\bar{\xi}^2+4\,\bar{\xi}\,Q_3+4\,\bar{\xi}^2\,Q_3+\right.$$
$$\left.10\,\bar{\xi}^2\,Q_4+6\,\bar{\xi}^3\,Q_4+12\,\bar{\xi}^3\,Q_5+9\,\bar{\xi}^4\,Q_6\right)\bar{N}^6\,\bar{\xi}(L), \quad (60)$$

$$\left(\Delta^{\rm finite}\tilde{F}_3\right)^2_{\rm BeS} = 9\left(1+2\,\bar{\xi}+\bar{\xi}^2+4\,\bar{\xi}\,Q_3+4\,\bar{\xi}^2\,Q_3+2\,\bar{\xi}^2\,{Q_3}^2-2\,\bar{\xi}^3\,{Q_3}^2-\right.$$
$$4\,\bar{\xi}^3\,{Q_3}^3+\bar{\xi}^4\,{Q_3}^4+8\,\bar{\xi}^2\,Q_4+8\,\bar{\xi}^3\,Q_4+16\,\bar{\xi}^3\,Q_3\,Q_4-$$
$$\left.8\,\bar{\xi}^4\,{Q_3}^2\,Q_4+16\,\bar{\xi}^4\,{Q_4}^2\right)\bar{N}^6\,\bar{\xi}(L), \quad (61)$$

$$\left(\Delta^{\rm edge}\tilde{F}_3\right)^2_{D=2} = \left(24.23+80.71\,\bar{\xi}+31.50\,\bar{\xi}^2+131.6\,\bar{\xi}\,Q_3+306.0\,\bar{\xi}^2\,Q_3+117.6\,\bar{\xi}^3\,{Q_3}^2+\right.$$
$$\left.504.0\,\bar{\xi}^2\,Q_4+409.6\,\bar{\xi}^3\,Q_4+1280\,\bar{\xi}^3\,Q_5+1805\,\bar{\xi}^4\,Q_6\right)\bar{N}^6\,\bar{\xi}\,\frac{v}{V}, \quad (62)$$

$$\left(\Delta^{\rm edge}\tilde{F}_3\right)^2_{D=3} = \left(34.62+99.26\,\bar{\xi}+39.60\,\bar{\xi}^2+180.3\,\bar{\xi}\,Q_3+331.1\,\bar{\xi}^2\,Q_3+93.50\,\bar{\xi}^3\,{Q_3}^2+\right.$$
$$\left.633.5\,\bar{\xi}^2\,Q_4+441.3\,\bar{\xi}^3\,Q_4+1379\,\bar{\xi}^3\,Q_5+1668\,\bar{\xi}^4\,Q_6\right)\bar{N}^6\,\bar{\xi}\,\frac{v}{V}, \quad (63)$$

$$\left(\Delta^{\rm discrete}\tilde{F}_3\right)^2_{D=2} = \left(1.593+8.273\,\bar{N}+9.\,\bar{N}^2+4.263\,\bar{\xi}+43.37\,\bar{N}\,\bar{\xi}+76.22\,\bar{N}^2\,\bar{\xi}+\right.$$
$$19.01\,\bar{N}\,\bar{\xi}^2+97.22\,\bar{N}^2\,\bar{\xi}^2+3.814\,\bar{\xi}^2\,Q_3+76.06\,\bar{N}\,\bar{\xi}^2\,Q_3+$$
$$194.4\,\bar{N}^2\,\bar{\xi}^2\,Q_3+166.0\,\bar{N}^2\,\bar{\xi}^3\,Q_3+89.22\,\bar{N}\,\bar{\xi}^3\,Q_4+442.7\,\bar{N}^2\,\bar{\xi}^3\,Q_4+$$
$$\left.592.9\,\bar{N}^2\,\bar{\xi}^4\,Q_5\right)\,\bar{N}^3\,\frac{v}{V}, \quad (64)$$

$$\left(\Delta^{\rm discrete}\tilde{F}_3\right)^2_{D=3} = \left(0.879+5.829\,\bar{N}+9.\,\bar{N}^2+2.116\,\bar{\xi}+27.13\,\bar{N}\,\bar{\xi}+66.53\,\bar{N}^2\,\bar{\xi}+\right.$$

$$10.59\,\bar{N}\,\bar{\xi}^2 + 74.23\,\bar{N}^2\,\bar{\xi}^2 + 1.709\,\bar{\xi}^2\,Q_3 + 42.37\,\bar{N}\,\bar{\xi}^2\,Q_3 +$$
$$148.5\,\bar{N}^2\,\bar{\xi}^2\,Q_3 + 111.2\,\bar{N}^2\,\bar{\xi}^3\,Q_3 + 44.40\,\bar{N}\,\bar{\xi}^3\,Q_4 + 296.4\,\bar{N}^2\,\bar{\xi}^3\,Q_4 +$$
$$349.3\,\bar{N}^2\,\bar{\xi}^4\,Q_5\Big)\;\bar{N}^3\,\frac{v}{V}. \tag{65}$$

Let us discuss the significance and meaning of the different terms in these equations. $\left(\Delta^{\rm finite}\tilde{F}_k\right)^2_{\rm X}$, with X=SS or BeS, is proportional to $\bar{\xi}(L)$ (and formally independent of the spatial dimension $D$). It corresponds to the *finite volume effect* due to the fluctuations of the underlying random field at scales larger than the sample size. The relative finite volume error $\Delta^{\rm finite}\tilde{F}_k/F_k$ is independent of $\bar{N}$, or, in other words, independent of the number of objects $N_{\rm par}$ in the catalog, in accordance with intuition. The two approximations SS and BeS, although quite different formally, gave similar result in all practical cases studied. To illustrate this point, we computed the expected finite volume errors in these two approximations for a three-dimensional hierarchical sample $\mathcal{S}$ with power law correlation function $\bar{\xi} = (\ell/\ell_0)^{-\gamma_S}$, $\gamma_S = 1.8$, and $\ell_0 = L/20$. For the amplitude of higher order correlations we take the $Q_N$'s measured by Gaztañaga (1994) in the APM survey (corrected to have three-dimensional statistics, see also Szapudi *et al.* 1995)

$$Q_3 = 1.35, \quad Q_4 = 2.33, \quad Q_5 = 4.02, \quad Q_6 = 6.7, \quad Q_7 = 10, \quad Q_8 = 12. \tag{66}$$

Figure 2 displays the quantity $\Delta^{\rm finite}\tilde{F}_k/F_k$ as a function of $\ell/L$. Each panel corresponds to a given order $k$. The difference between the two approximations is mostly negligible, at most 20%. $\Delta^{\rm finite}\tilde{F}_k/F_k$ increases with $k$, as expected, and it exhibits two plateaux: one in the weakly nonlinear regime $\bar{\xi}_2 \ll 1$, and one in the highly nonlinear regime $\bar{\xi}_2 \gg 1$, as can be easily inferred from equations (50), (54), (55), (60) and (61).

The term $\left(\Delta^{\rm edge}\tilde{F}_k\right)^2$ corresponds to the *edge effects.* It is due to the fact that the statistical weight given to objects near the edge is smaller than far away from it. It can be clearly disentangled from the *discreteness effect* term $\left(\Delta^{\rm discrete}\tilde{F}_k\right)^2$ corresponding to the "shot noise" introduced by the finite number of objects in the catalog. Indeed, in the continuous limit ($\bar{N} \rightarrow \infty$) the relative error $\Delta^{\rm discrete}\tilde{F}_k/F_k$ vanishes, whereas the relative error $\Delta^{\rm edge}\tilde{F}_k/F_k$ is independent on $\bar{N}$, and proportional to $\sqrt{\bar{\xi}v/V}$ up to leading order in $v/V$. This also suggests that edge effects are non-existent (or very weak) for a Poisson (or a weakly clustered) sample, confirming the intuition that geometry is unimportant for (nearly) Poisson statistics. To illustrate these points, Figure 3 displays the quantities $\Delta^{\rm edge}\tilde{F}_k/F_k$ and $\Delta^{\rm discrete}\tilde{F}_k/F_k$ for our reference catalog $\mathcal{S}$, assuming various values of $N_{\rm par} = 500$, 5000, 50000. As expected, discreteness effects are larger at smaller scales, particularly when the number of objects $N_{\rm par}$ is small, while edge effects increase with $\ell$.

The formal expression for the edge and discreteness effects is

$$\left(\Delta^{\rm X}\tilde{F}_k\right)^2 = \sum c_{h,i,j_1,\ldots,j_{2N}}\bar{N}^h\bar{\xi}^i Q_1^{j_1}\ldots Q_{2N}^{j_{2N}}\frac{v}{V}, \tag{67}$$

with X=edge or X=discrete. The numerical value of the coefficients $c_{h,i,j_1,\ldots,j_{2N}}$ is fairly insensitive to changes of $\gamma_3$ in the regime consistent with the observations (moreover the two and three-dimensional coefficients are quite similar), therefore the quoted equations constitute good approximations even for $\gamma_3 \neq 1.8$. To show this, we computed $c_{h,i,j_1,\ldots,j_{2N}}$ for $D = 3$, $k \leq 4$, and for various values of $\gamma = 1.8$, 1.5, 1.2 and 0.9. The corresponding errors $\left(\Delta^{\rm overlap}\tilde{F}_k/F_k\right)^2 =$

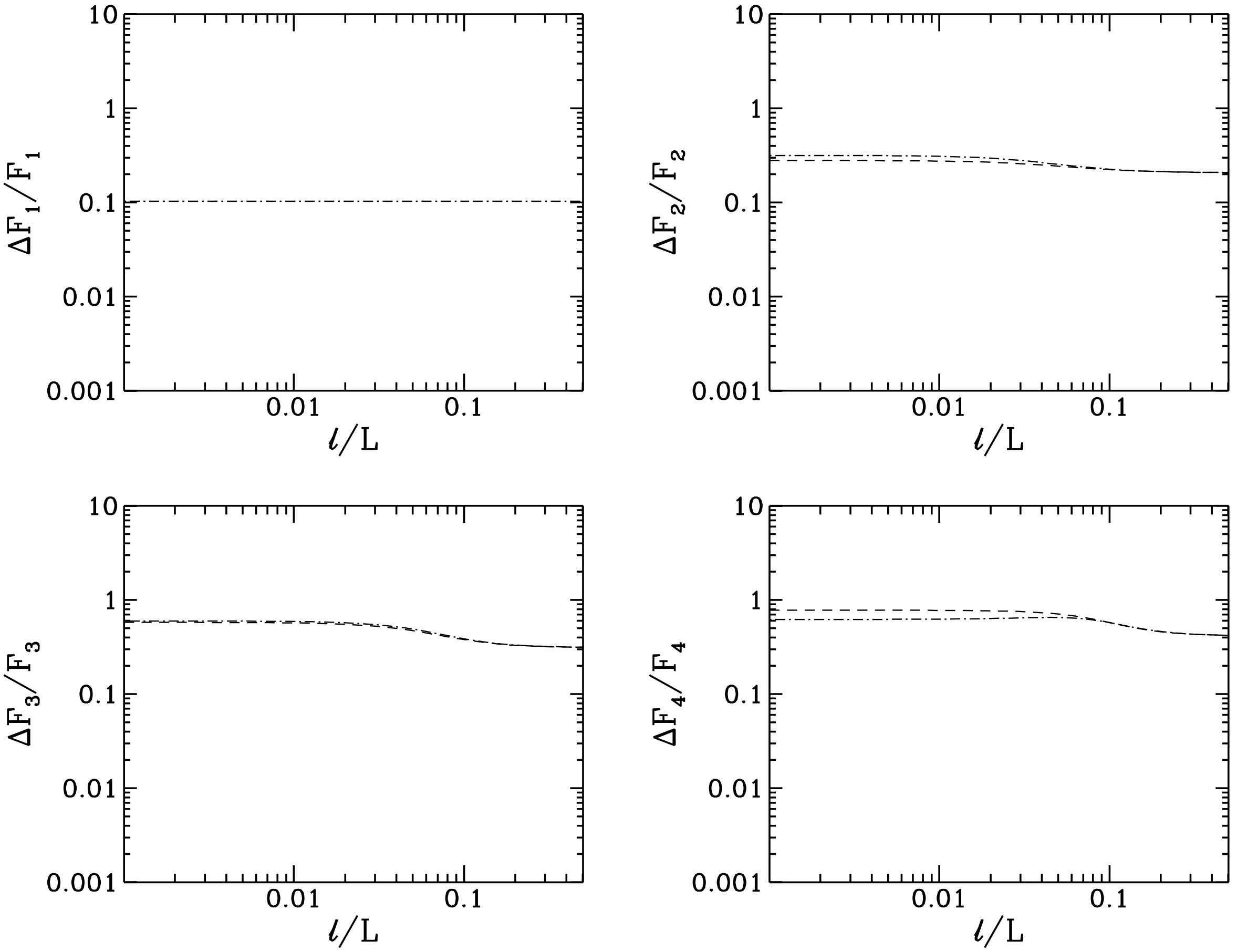

Fig. 2.— The relative finite volume error $\Delta^{\rm finite}\tilde{F}_k/F_k$ brought by fluctuations of the underlying random field at scales larger than the sample size $L \sim V^{1/3}$ is plotted as a function of cell size $\ell$, for our reference catalog $\mathcal{S}$ (see text). Each panel corresponds to a given value of $k$. The dotted-dashed and dashed curves correspond respectively to the approximations SS and BeS discussed in § 3. For $k = 1$, both approximations give the same results. For $k \geq 2$, they differ only by a small amount.

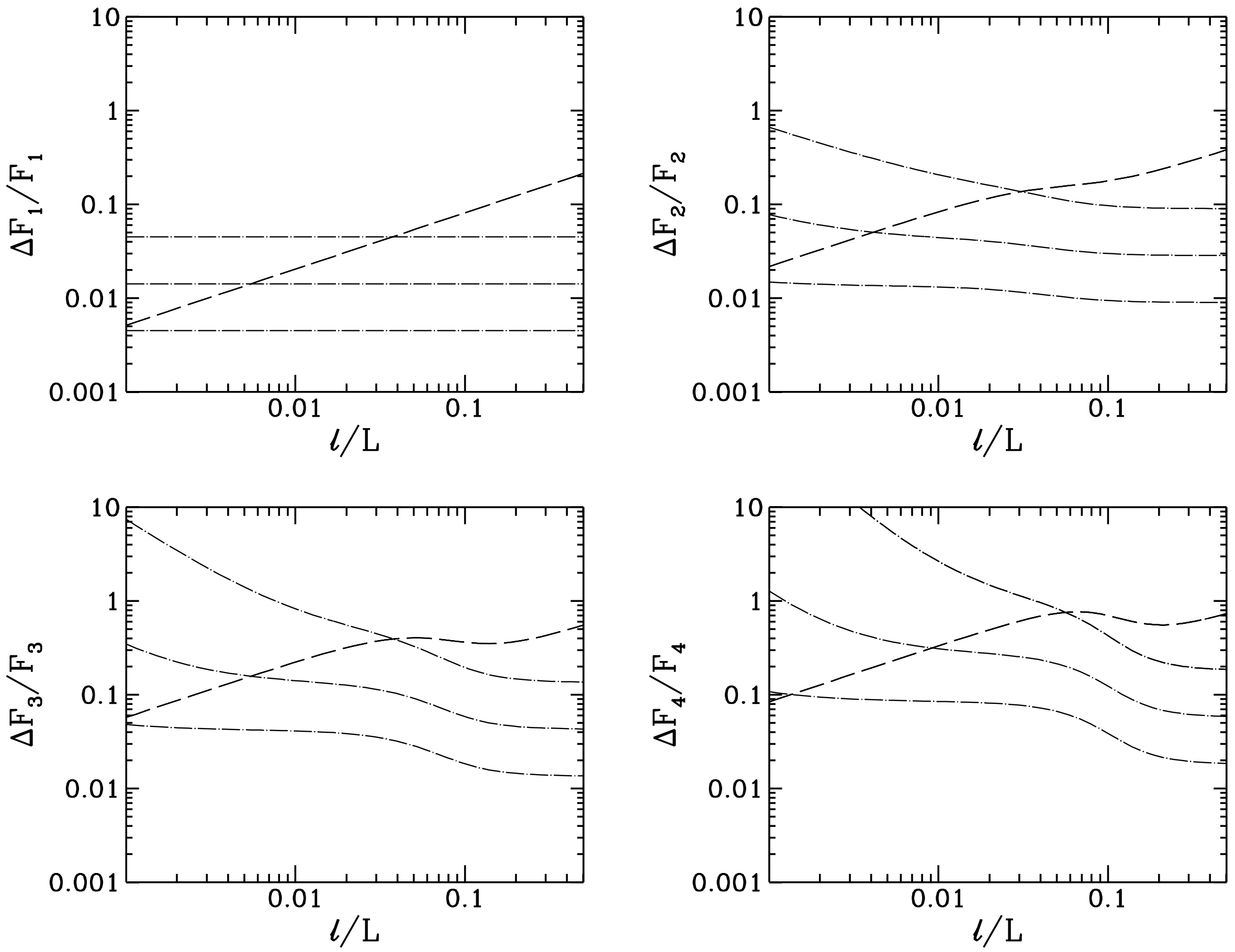

Fig. 3.— The error $\Delta^{\rm edge}\tilde{F}_k/F_k$ brought by edge effects (long dashes) and the error $\Delta^{\rm discrete}\tilde{F}_k/F_k$ due to the finite number of objects in the catalog (dots-long dashes) are displayed as functions of cell size $\ell$, for our reference set $\mathcal{S}$ (see text). The number of objects in $\mathcal{S}$ is assumed to be $N_{\rm par}$ = 500, 5000 and 50000. When $N_{\rm par}$ increases, $\Delta^{\rm discrete}\tilde{F}_k/F_k$ decreases whereas $\Delta^{\rm edge}\tilde{F}_k/F_k$ remains constant.

$\left(\Delta^{\rm edge}\tilde{F}_k/F_k\right)^2 + \left(\Delta^{\rm discrete}\tilde{F}_k/F_k\right)^2$ for our reference set $\mathcal{S}$, assuming that $N_{\rm par} = 5000$, are displayed on Figure 4 as functions of $\ell/L$. Each panel corresponds to a given value of the order $k$. Note that on Figure 4, the slope of the correlation function, $\gamma_{\mathcal{S}}$, used to evaluate $\bar{\xi}$, is held fixed.

## 5. Example: a Rayleigh-Lévy Fractal

In the previous section, two important hypotheses made the calculation of the errors possible: hierarchical model and locally Poisson behavior. The first assumption is supported by the statistical properties of the galaxy distribution and the measurements in $N$-body simulations. Here, to justify the second ansatz, we estimate experimentally the errors on measurements of factorial moments in subsamples of a homogeneous Rayleigh-Lévy fractal described in § 5.1, and compare the results to our theoretical predictions in § 5.2. Indeed, such a control sample is strongly clustered, thus far from showing a locally Poisson behavior. Another significant point, discussed in the introduction, is to examine the distribution of errors rather than just the dispersion of the measurements. Since we have access only to a unique part of the Universe, it is important to know to what extent the cosmic errors are systematic. To study this question analytically with the same degree of generality we used until now would be rather difficult. Instead, we measure in § 5.3 the distribution of errors in our control sample.

### 5.1. The sample

The sample $\mathcal{F}$, of $N_{\rm par} = 128^3$ points, was generated in a three dimensional unit torus (a cube of size $L_{\mathcal{F}} \equiv 1$ with periodic boundary conditions) using 1024 Rayleigh-Lévy random walks $\mathcal{W}_i \equiv \{\mathcal{W}_{i,j},\ j = 1, 2048\}$. Each walk starts from a point $\mathcal{W}_{i,1}$ at a random position. In a given walk $\mathcal{W}_i$, the next point $\mathcal{W}_{i,j+1}$ is chosen at random direction and at distance $r$ from $\mathcal{W}_{i,j}$ drawn from the following distribution

$$\begin{array}{ll} p(r > \ell) = (\ell_{\rm p}/\ell)^{\epsilon}, & \ell \geq \ell_{\rm p}, \\ p(r > \ell) = 1, & \ell < \ell_{\rm p}. \end{array} \tag{68}$$

The statistical properties of a Rayleigh-Lévy fractal can be fully calculated once $N_{\rm par}$, $\epsilon$, and the percolation length $\ell_{\rm p}$ are known, in particular, the two-point correlation function $\xi(r)$ (Mandelbrot 1975; Peebles 1980), hence $\bar{\xi}$ (see details in CBSII). Here we chose $\ell_{\rm p} = 4.38\,10^{-4}$ and $\epsilon = 1.2$ so that $\bar{\xi} = \left(\frac{\ell}{\ell_0}\right)^{-\gamma_{\mathcal{F}}}$ with $\ell_0 = L_{\mathcal{F}}/40$ and $\gamma_{\mathcal{F}} = 1.8 = 3 - \epsilon$. It can be easily shown that such a fractal obeys the special hierarchical model of equation (22) (see, e.g., Hamilton & Gott 1988; Bernardeau & Schaeffer 1992; CBSII; and eq. [30]). The approximation BeS for the finite volume effect error is thus exact in this particular case (up to leading order in $v/V$). To compute the errors on $F_k$ up to $k = 4$, we need the values of $Q_N$ up to $N = 8$. We have approximately $Q_N \simeq 2^{1-N} N!/N^{N-2}$ with some small correction from the effect of the smoothing over the cell. Following CBSII (see their Table 1), we have, with this correction,

$$Q_3 \simeq 0.514, \quad Q_4 \simeq 0.200, \quad Q_5 \simeq 0.0662, \quad Q_6 \simeq 0.0199, \quad Q_7 \simeq 0.0056, \quad Q_8 \simeq 0.0015. \tag{69}$$

Figure 5 displays a thin slice of $\mathcal{F}$ ($L_{\mathcal{F}}/50$ thick), showing the clumpy nature of our fractal, which is thus far from being locally Poisson.

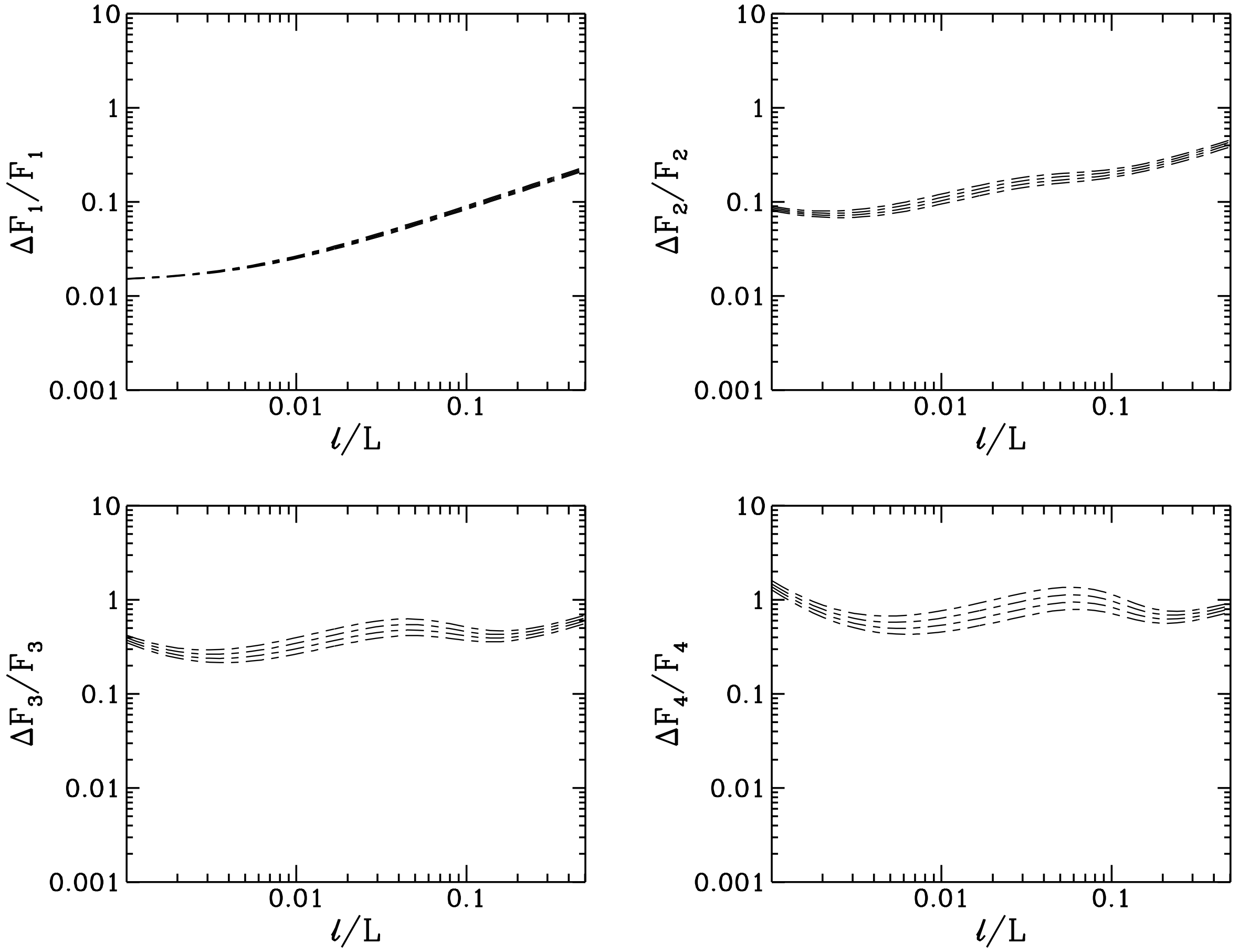

Fig. 4.— This figure displays the dependence of the "overlapping" error, $\left(\Delta^{\rm overlap}\tilde{F}_k/F_k\right)^2 = \left(\Delta^{\rm edge}\tilde{F}_k/F_k\right)^2 + \left(\Delta^{\rm discrete}\tilde{F}_k/F_k\right)^2$, on the value of $\gamma$ used to compute the coefficients $c_{h,i,j_1,\ldots,j_{2N}}$ in equation (67). The quantity $\Delta^{\rm overlap}\tilde{F}_k/F_k$ is plotted as a function of cell size, for $\gamma = 1.8$, 1.5, 1.2 and 0.9 and for our reference sample $\mathcal{S}$ (without changing $\gamma_{\mathcal{S}}$), assuming that it contains $N_{\rm par} = 5000$ objects. The value of $\Delta^{\rm overlap}\tilde{F}_k/F_k$ is increasing with $1/\gamma$.

Fig. 5.— A thin slice ($L_{\mathcal{F}}/50$ thick) of our Rayleigh-Lévy fractal $\mathcal{F}$ is shown (see text).

### 5.2. Comparison of the Theoretical Predictions with Numerical Results

From $\mathcal{F}$, we extracted $N_{\rm sub} = 1000$ cubical subsamples $\mathcal{F}^i_{\rm sub}$ of size $L = L_{\mathcal{F}}/4$. The position of each subsample was choosen randomly. We also randomly diluted the subsamples by a factor $\alpha^{-1} = 64$. In other words, the probability that a particle in the volume intersecting $\mathcal{F}^i_{\rm sub}$ was selected is $p = \alpha$. In the following, if $A$ is a statistical quantity, $\tilde{A}^i$ is its measurement in $\mathcal{F}^i_{\rm sub}$, and $\left\langle \tilde{A} \right\rangle_i \equiv N_{\rm sub}^{-1} \sum_i \tilde{A}_i$ is the average of $\tilde{A}^i$ over the $N_{\rm sub}$ realizations. If $\tilde{N}^i_{\rm par}$ is the number of particles per subsample, its average is thus $\left\langle \tilde{N}_{\rm par} \right\rangle_i \simeq \alpha N_{\rm par} = 512$. In each subsample we measured the factorial moments $\tilde{F}^i_k$, using a large number $C \geq 128^3$ of cells so that the measurement error $\Delta^{C,\infty} \tilde{F}_k$ was negligible. The (biased) experimental estimate of the error on $\tilde{F}^i_k$ is thus

$$\left(\Delta \tilde{F}_k\right)^2 = \left\langle \left(\delta \tilde{F}_k\right)^2 \right\rangle_i \tag{70}$$

with

$$\delta \tilde{F}^i_k = \tilde{F}^i_k - \left\langle \tilde{F}_k \right\rangle_i . \tag{71}$$

Similarly to the previous considerations, there is *measurement error on the error* due the finite number of subsamples extracted from $\mathcal{F}$, and there is *cosmic error on the error* due to the finite size of $\mathcal{F}$, its geometry and the finite number of objects it contains. The former can be estimated by straightforward error propagation, the latter is rather complicated because it depends on, e.g., $16^{\rm th}$ order quantities for the $4^{\rm th}$ order cosmic error on the error.

Figure 6 displays the measured value of $\Delta \tilde{F}_k / F_k$ (circles) versus the theoretical predictions (solid curves), which use the approximation BeS to compute the finite volume error $\Delta^{\rm finite} \tilde{F}_k / F_k$ (short dashes). Note that the SS approximation would give similar results. The edge effect contribution $\Delta^{\rm edge} \tilde{F}_k / F_k$ (long dashes) and the discreteness contribution $\Delta^{\rm discrete} \tilde{F}_k / F_k$ (dot-long dashes) are also displayed. There are errorbars on the dots, when sufficiently large to be visible. They correspond to the measurement error on the error. Note the *excellent agreement of our predictions with the numerical experiment*, which also shows that the cosmic error on the error problem mentioned above is well controlled in our experiment. This particular example also illustrates how the shot noise contribution $\Delta^{\rm discrete}\tilde{F}_k/F_k$ to the error increases with $k$ at small scales. The edge effect contribution is negligible at small $\ell$ but always dominates at large scales, although the steadier finite volume effect contribution is of the same order in this regime.

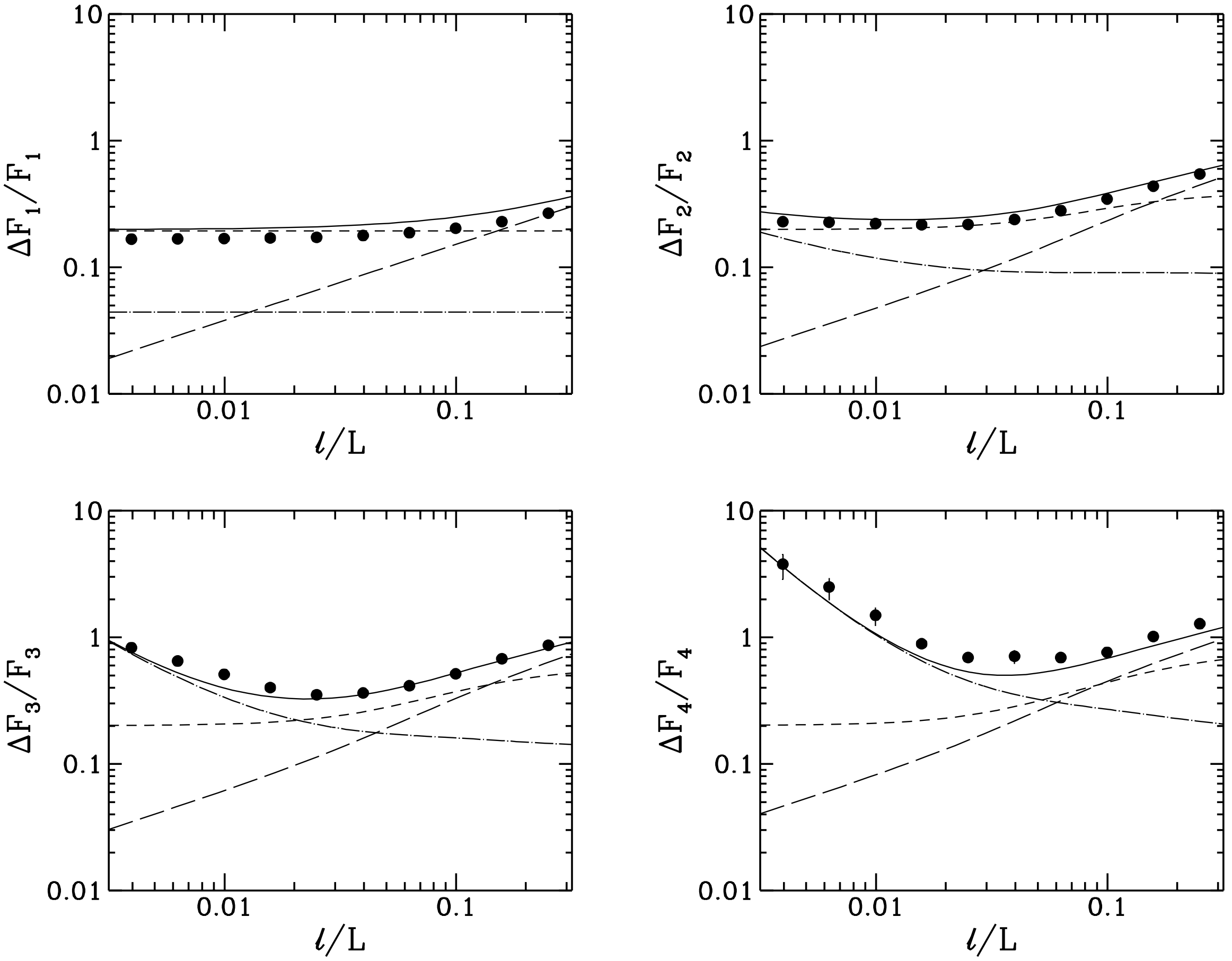

Fig. 6.— This figure shows the measured value of the error $\Delta\tilde{F}_k/F_k$ on the measurement of $F_k$ in the fractal subsamples $\mathcal{F}^i_{\rm sub}$ as a function of scale (dots), and our theoretical prediction (continuous curve). Each panel corresponds to a value of $k$. The BeS approximation was used to compute the finite volume error $\Delta^{\rm finite}\tilde{F}_k/F_k$ (short dashes). The edge effect contribution $\Delta^{\rm edge}\tilde{F}_k/F_k$ (long dashes) and the discreteness contribution $\Delta^{\rm discrete}\tilde{F}_k/F_k$ (dot-long dashes) are also displayed.

### 5.3. Distribution of the Errors

Let us imagine now that we have access to only one subsample $\mathcal{F}^i_{\rm sub}$ and measure $\tilde{F}^i_k$. Is it more likely to under- or overestimate the real value of $F_k$? To answer this question, we measured the probability distribution function $\Upsilon(\delta\tilde{F}^i_k)$ of the errors. The result is displayed in Figure 7 for two scales, $L/64$ (four top panels, each one corresponding to a given value of $k$) and $L/4$ (four bottom panels). On each panel, the errorbars on the measurements reflect the finite number of subsamples, and the continuous curve corresponds to a Gaussian with average zero and variance $\left(\Delta\tilde{F}_k\right)^2$. From Figure 7, the distribution of the errors is increasingly skewed with both increasing $k$ and scale compared to a Gaussian. In particular, the maximum of function $\Upsilon(\delta\tilde{F}^i_k)$ lies at a *negative* value of $\delta F^i_k$, which indicates that it is likely to *underestimate* the real value of $F_k$, especially if the order $k$ and/or the cell size $\ell$ is large. This systematic effect was already pointed out by CBSI and CBSII who proposed a method to correct for it in some particular regimes using some assumptions on the asymptotic properties of the probability distribution.

## 6. Discussion

In this paper we calculated the theoretical error on statistics related to counts in cells in a finite galaxy catalog. We identified the different contributions to the total theoretical error (the systematics of the observations are disregarded): the *measurement error*, due to the finite number of cells thrown to estimate the count probabilities (this in principle can be eliminated with the algorithm of Szapudi 1995), and the *cosmic error*, inherent to any finite catalog. The cosmic error itself has three contributions: the *discreteness effect* arising from sampling the underlying continuous random field with finite number of points, the *edge effect* caused by the lesser statistical weight given to objects near the boundary of the survey, and the *finite volume effect*, from fluctuations of the underlying random field at scales larger than the sample size. First, we presented the general mathematical formulation of the problem establishing a firm groundwork for subsequent applications, and solving the practical measurement error problem. For the cosmic error, in the framework of the hierarchical tree assumption, we have found a good approximation using a locally Poisson ansatz. The results, in excellent agreement with our control measurements performed on a Rayleigh-Lévy hierarchical sample, give a simple and useful way of estimating the expected errors for a galaxy survey with prescribed properties. Measurements of the distribution of the errors on the same Rayleigh-Lévy fractal showed that the cosmic errors tend to be systematic, i.e., it is more likely to *underestimate* the true value of $F_k$ than to overestimate it, as already pointed out by CBSI.

To further illustrate our results, we discuss here two important subjects. The first one concerns the dependence of the errors on the clustering properties of the underlying distribution. In partic-ular, is it fair to assume Gaussian behavior to compute the errors on statistics related to counts in cells? The second subject is the previously mentioned (§ 2) concept of "number of statistically independent cells". Further applications, such as reanalyses of the cosmic error on counts in cells measured in galaxy catalogs, or sampling strategies for galaxy surveys (Kaiser 1986) will be discussed elsewhere (Colombi, Szapudi & Szalay 1995). Note also, as mentioned in § 3, that the hierarchical model is not expected to be valid in the weakly nonlinear regime, although it should be a good approximation for the calculation of the errors (Bernardeau 1994a). More detailed investigations of the weakly nonlinear regime are left for future work.

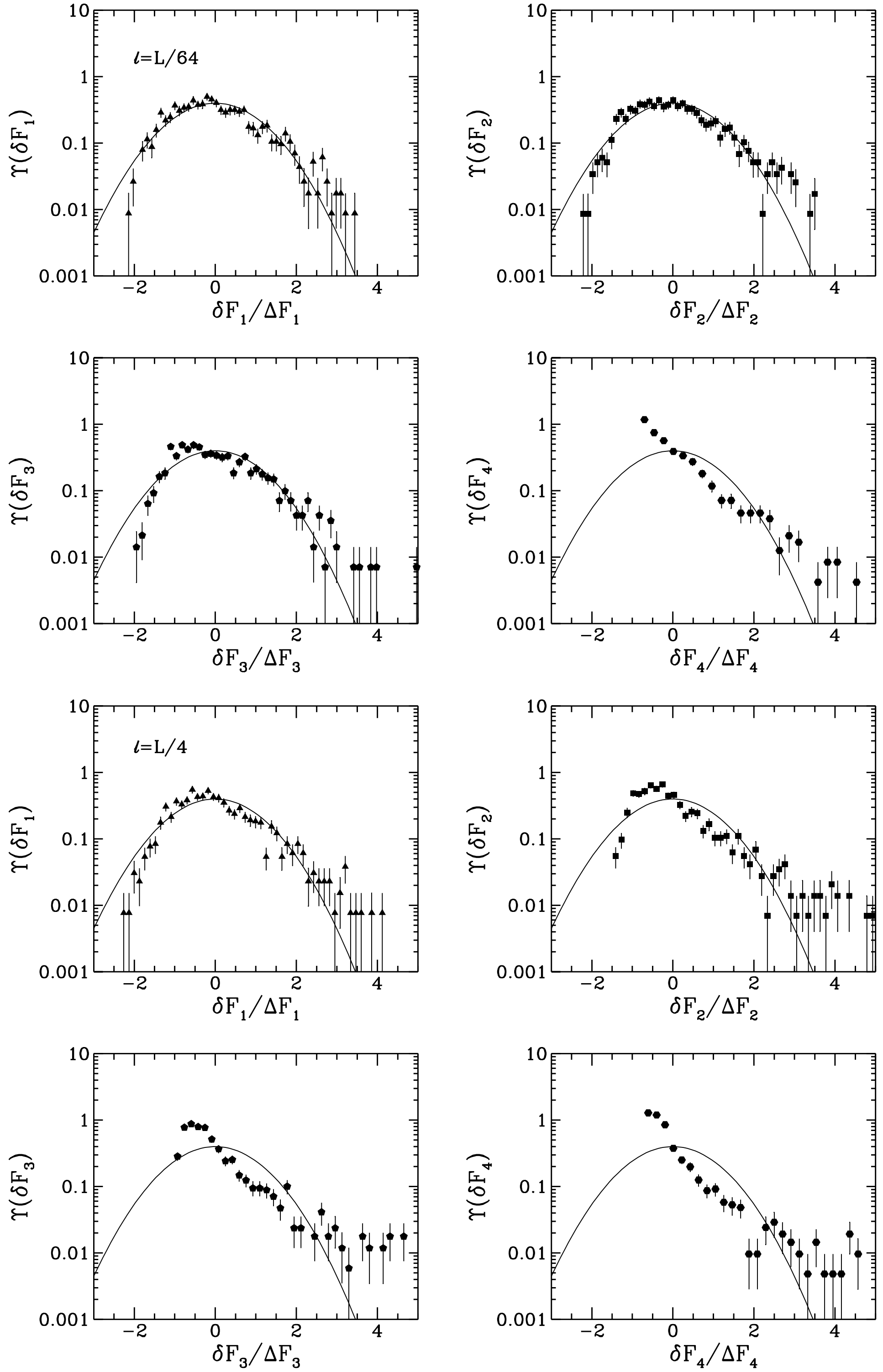

Fig. 7.— The probability distribution function $\Upsilon(\delta\tilde{F}_k^i)$ of the errors as a function of $\delta\tilde{F}_k^i/\Delta\tilde{F}_k$, measured in our 1000 subsamples $\mathcal{F}_{\rm sub}^i$. The four upper panels, each one for a given $k$, correspond to cell size $\ell = L/64$. The four lower panels correspond to $\ell = L/4$. On each panel, a Gaussian with average zero and with variance $\left(\Delta\tilde{F}_k\right)^2$ is displayed.

In what follows, the approximation SS discussed in § 3 was used to compute the finite volume contribution to the error.

Figure 8 illustrates the difference between Gaussian and full hierarchical assumptions for error calculations by displaying the expected cosmic error on the measurement of $F_k$ for samples of volume $V = L^3$, correlation function $\bar{\xi} = (\ell/\ell_0)^{-\gamma}$ with $\gamma = 1.8$ and $\ell_0 = L/20$. All these fictive samples have the same number of objects $N_{\rm par} = 5000$, however, the higher order statistics is varied. Each panel corresponds to a given value of $k$. The long dashes assume Gaussian underlying statistics for the calculation of $F_k$ and the calculation of the error. For $k \geq 2$, The (upper) short dashes, dots, and continuous curve assume that the hierarchy of $Q_N$'s is given by perturbation theory predictions (see, e.g., Bernardeau 1994b) for an initial power spectrum $\langle|\delta_k|^2\rangle \propto k^n$ with $n = -2, -1, 0$, respectively. For $k \geq 3$, the lower short dashes, dots and continuous curve assume the same hierarchy of $Q_N$'s for the calculation of $F_k$, but the errors are computed from Gaussian statistics. In the case $k = 2$, such an assumption would lead to the long dashes, whatever the values of the $Q_N$'s. In the case $k = 1$, the error depends only on statistics up to second order, for which all the models under consideration are equivalent. These plots clearly show a strong dependence of the cosmic error on the underlying clustering properties of the system. Moreover, assuming Gaussian statistics to compute the errors seems unreasonable: the errors can be severely underestimated, except, as expected, in weakly nonlinear regime.

Let us turn to a widely used but seldom explained concept: the number of statistically independent cells. We defined it in § 2 as the number of cells $C^*$ needed to sample the catalog so that the measurement error equals the cosmic error. This definition ensures that most of the statistically relevant information is extracted. In that sense, these cells can be considered as "statistically independent". However, there must remain residual information in the survey obtainable via more sampling cells, since the overall error can be decreased by another factor of two. This illustrates the level of arbitrariness in the concept of "number of statistically independent cells" for a sample of finite volume. Another popular but certainly erroneous choice is $C^* = C_V = V/v$, the number of cells needed to cover the sample volume $V$. To compare this choice with our more natural definition, Figure 9 displays the quantity $C^*/C_V$ for our reference sample $\mathcal{S}$ (see § 4.3), assuming that it contains $N_{\rm par} = 5000$ objects (left panel) and in the continuous limit $N_{\rm par} = \infty$ (right panel). It has been computed from the errors on $F_k$, with $k = 1$ (solid curves), 2 (dots), 3 (short dashes) and 4 (long dashes). First thing to notice is that the number of "statistically independent cells" is not universal: $C^*$ depends on the statistical object under study. In our example it increases with $k$. It is generally different from $C_V$ by several orders of magnitude. Note that $C^*$ is smaller in the continuous limit than for finite $N_{\rm par}$, contrarily to the expectation motivated by the fact that the shot noise tends to increase the cosmic errors. Another shot noise contribution to the measurement error, however, increases when $N_{\rm par}$ is small, thus explaining this counterintuitive effect. As already discussed in § 2, throwing a number of sampling cells $C \gg C^*$ would decrease the overall errors by a factor two. Moreover $C^*$ highly depends on the statistical object under study, so we endorse the use of *as many cells as possible* for counts in cells measurements, or to use an algorithm which is equivalent to throwing infinite number of cells (Szapudi 1995).

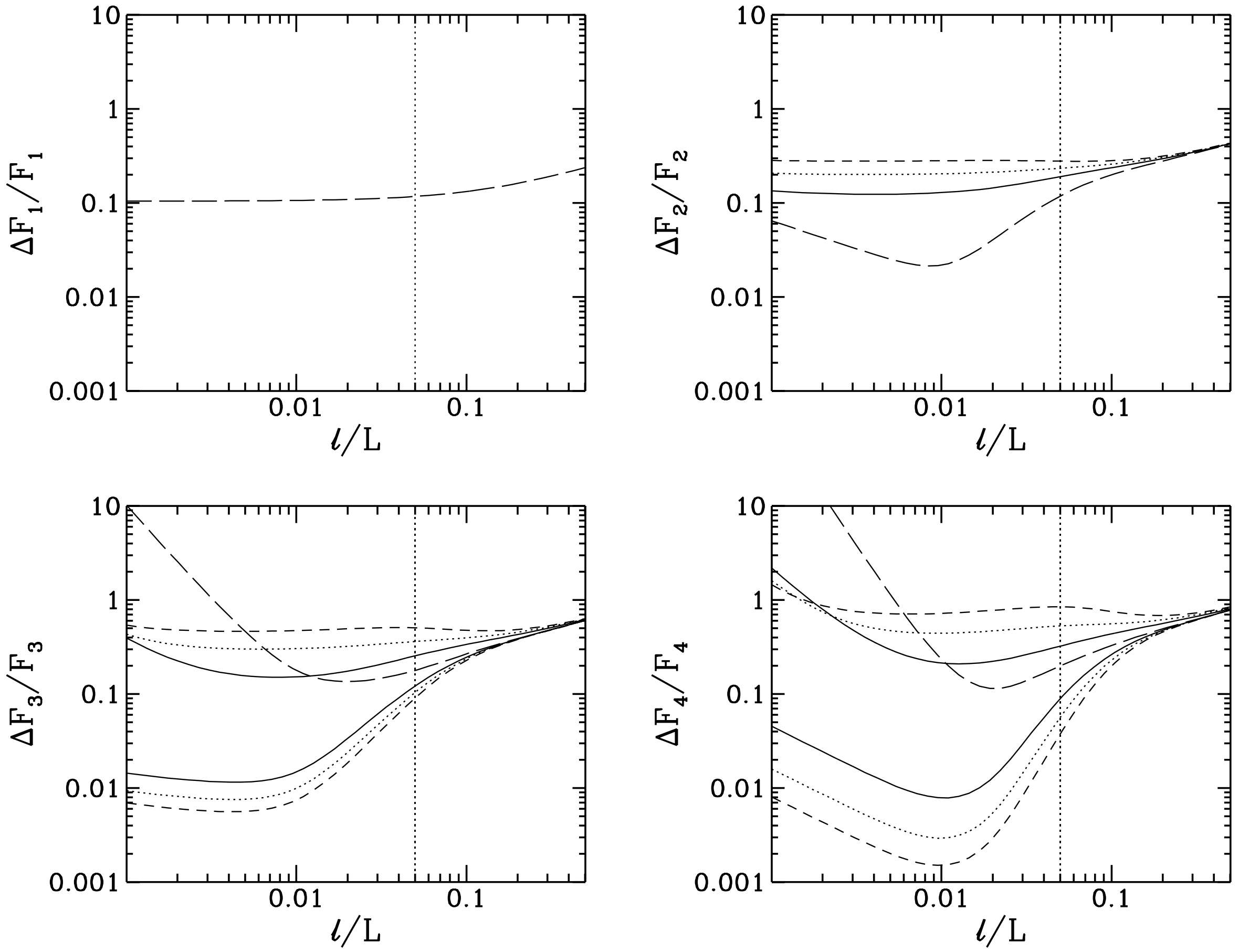

Fig. 8.— This figure displays the relative cosmic error on the measured factorial moment $F_k$ ($k = 1, \ldots, 4$), as a function of the cell size $\ell$ for a catalog of volume $V = L^3$, assuming that it contains $N_{\rm par} = 5000$ objects and its correlation function is $\bar{\xi} = (\ell/\ell_0)^{-1.8}$ with $\ell_0 = L/20$. Each panel corresponds to a different value of $k$. The long dashes assume underlying Gaussian statistics. For $k \geq 2$, the upper dots, short dashes and continuous curves assume that higher order statistics is given by perturbation theory predictions for scale invariant initial power-spectra of indexes $n = -2$, -1, 0 respectively. For $k \geq 3$, there are lower dots, short dashes and continuous curve. In that case, that the error $\Delta \tilde{F}_k$ has been calculated assuming Gaussian statistics, but the $F_k$ remain unchanged. The vertical dotted line on each panel marks the value of the correlation length $\ell_0$.

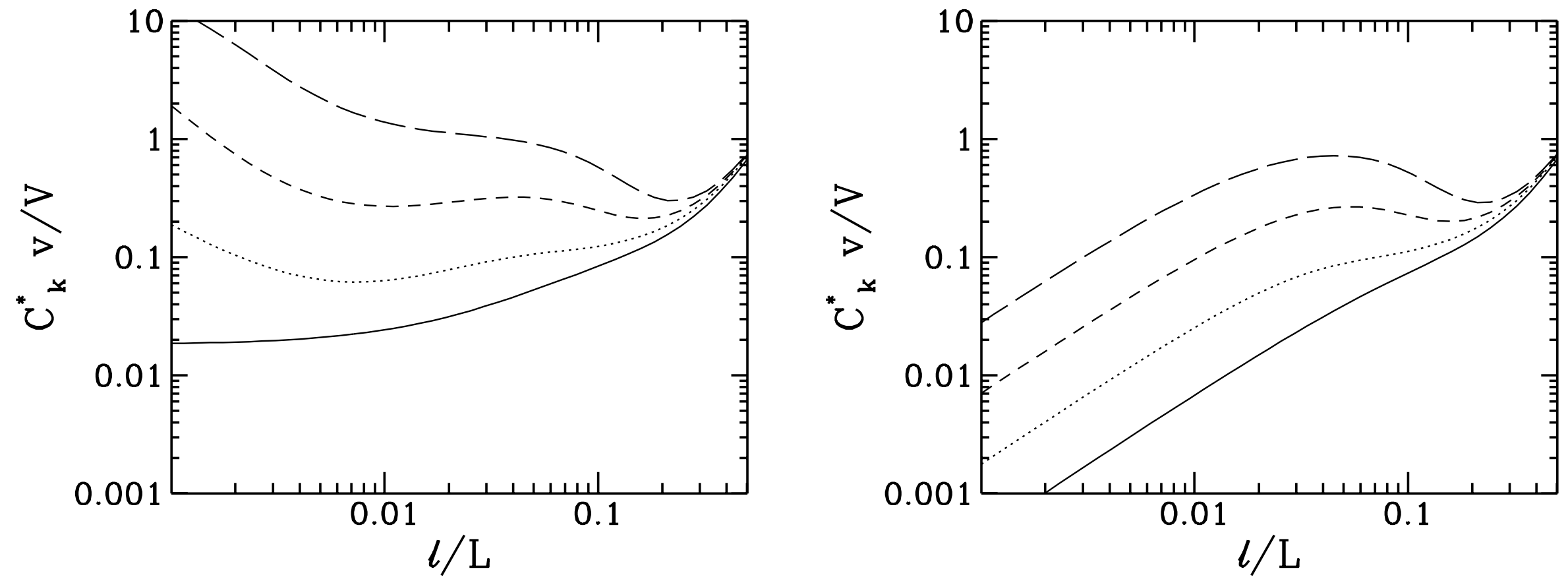

Fig. 9.— The "number of statistically independent cells" $C^*$ for the factorial moments of order $k$. It is divided by the number of cells needed to cover the catalog $C_V = V/v$ and plotted as a function of cell size for our reference catalog $\mathcal{S}$. The left panel assumes that it contains $N_{\rm par} = 5000$ objects, while the right panel corresponds to the continuous limit. The solid curve, dots, short dashes and long dashes correspond respectively to $k = 1$, 2, 3 and 4.

We would like to thank Francis Bernardeau, François Bouchet, Richard Schaeffer and Alex Szalay for stimulating discussions. This research was supported in part by DOE and by NASA through grant NAG-5-2788 at Fermilab. IS would like to thank the Aspen Center for Physics for its hospitality while some of this work was done.

## Appendix

### A. Generating Function for Overlapping Cells

In this Appendix, we show how to relate the bivariate generating function of overlapping cells to a trivariate generating function. Two overlapping cells can be imagined as three non-overlapping (touching) cells (see Fig. 1). Knowing the trivariate probability distribution $P_{H,I,J}$ of these three cells, it is simple to express the bivariate probability distribution for the two original cells

$$P_{M,N} = \sum_{H,I,J} \delta(H + I = N)\delta(I + J = M)P_{H,I,J}, \tag{A1}$$

where $I$ is the number count in the overlap area. The bivariate generating function can be expressed in terms of the trivariate generating function

$$P(x, y) = \sum_{M,N} P_{M,N} x^N y^M = P(x, xy, y), \tag{A2}$$

where

$$P(x,y,z) = \sum_{H,I,J} P_{H,I,J} x^H y^I z^J, \tag{A3}$$

is the generating function corresponding to the tree cells. Alternatively, the path integral formalism of SSI (see their eq. [5.5]) gives the same result with the following special source (see also BS)

$$\begin{aligned} J^*(x) &= W_{1\backslash 2,R}(x) z_1 + W_{1\cap 2,R}(x) z_1 z_2 + W_{2\backslash 1,R}(x) z_2, \\ W_{i,R}(x) &= \begin{cases} 1 \text{ if } x \in V_i \\ 0 \text{ if } x \notin V_i \end{cases}, \end{aligned} \tag{A4}$$

where the $W$'s are the characteristic functions of the union and the set theoretical differences of the two cells.

The generalization of the above formulae for $N$-variate generating functions with possible overlaps is a trivial, although tedious exercise.

## B. Contribution of the Error from Overlapping Cells

In this Appendix, we improve the approximation of equation (47) in three dimensions ($D = 3$). For simplicity, we assume that the survey is spherical of radius $R$ and that the origin of the coordinates is the center of the survey. The possible positions of cells of radius $\ell$ contained in the survey are thus $r \le \hat{R} \equiv R - \ell$. Introducing

$$\hat{\psi} \equiv \hat{R}/\ell, \quad \psi_1 \equiv r_1/\ell, \quad \psi \equiv |r_1 - r_2|/\ell, \tag{B5}$$

and using spherical coordinates, we have,

$$\begin{aligned} \left\langle \tilde{P}(x)\tilde{P}(y) \right\rangle_{\text{overlap}} &= \frac{v^2}{\hat{V}^2} \int_{0 \le \psi_1 \le \hat{\psi}} 3\psi_1^2 d\psi_1 \\ & \left\{ \int_{0 \le \psi \le \min(\hat{\psi}-\psi_1, 2)} 3\psi^2 d\psi + \right. \\ & \left. \int_{\hat{\psi}-\psi_1 \le \psi \le \min(\hat{\psi}, 2)} \left[ \frac{\hat{\psi}^2 - (\psi - \psi_1)^2}{4\psi\psi_1} \right] 3\psi^2 d\psi \right\} P^b(x,y). \end{aligned} \tag{B6}$$

This two-dimensional integral could be easily evaluated numerically, in case more than leading order accuracy in $v/V$ is needed to estimate the errors, particularly edge effects.

For the sake of comparison, we compute explicitly the dispersion on the average count using the above (more accurate) expression.

$$\left\langle \bar{N}^2 \right\rangle_{\text{overlap}} = \left\langle F_1^2 \right\rangle_{\text{overlap}} = \frac{\partial}{\partial x} \frac{\partial}{\partial y} \left\langle \tilde{P}(x+1)\tilde{P}(y+1) \right\rangle_{x=y=0}, \tag{B7}$$

where we dropped the obvious "overlap" index on the right hand side. We consider the particular case $V \ge 27v$ ($\hat{\psi} \ge 2$) and $\gamma = 3/2$. The result is

$$\left\langle \bar{N}^2 \right\rangle_{\text{overlap}} \simeq \frac{v}{\hat{V}} \left\{ \left[ \frac{2}{21}\frac{v}{\hat{V}} - \frac{27}{35}\left(\frac{v}{\hat{V}}\right)^{1/3} + 1 \right] \bar{N} \right.$$

$$+\left[2\frac{v}{\hat{V}} - 9\left(\frac{v}{\hat{V}}\right)^{1/3} + 8\right]\bar{N}^2$$
$$+\left[1.432\frac{v}{\hat{V}} - 6.516\left(\frac{v}{\hat{V}}\right)^{1/3} + 5.860\right]\bar{N}^2\bar{\xi}\Bigg\}. \tag{B8}$$

This equation is *exact* for a Poisson sample if we set $\bar{\xi} \equiv 0$. The approximation of equation (47) leads to

$$\left\langle \bar{N}^2 \right\rangle_{\rm overlap} \simeq \frac{v}{V}\left[\bar{N} + 8\bar{N}^2 + 5.860\bar{N}^2\bar{\xi}\right], \tag{B9}$$

showing that it is indeed correct up to the leading order in $v/V$. Note that for $v/V = 1/27$, equation (B8) gives

$$\left\langle \bar{N}^2 \right\rangle_{\rm overlap} = 0.078\bar{N} + 0.469\bar{N}^2 + 0.348\bar{N}^2\bar{\xi}, \tag{B10}$$

while under the same assumption (B9) yields

$$\left\langle \bar{N}^2 \right\rangle_{\rm overlap} \simeq 0.037\bar{N} + 0.296\bar{N}^2 + 0.217\bar{N}^2\bar{\xi}. \tag{B11}$$

The difference is less than a factor of 2, showing that the leading order in $v/V$ is still a reasonable approximation.